\documentclass[journal,onecolumn,11pt]{IEEEtran}
\ifCLASSINFOpdf
\else
\fi
\newcommand{\F}{\mathbb{F}}

\newcommand {\ccc}{{\mathbf{c}}}
\newcommand {\xxx}{{\mathbf{x}}}
\newcommand {\yyy}{{\mathbf{y}}}
\newcommand{\C}{{\mathcal{C}}}

\newcommand{\SSS}{{\mathcal{S}}}
\newcommand{\GRS}{{\mathrm{GRS}}}
\newcommand{\ESGRS}{{\mathrm{ESGRS}}}
\newcommand{\EGRS}{{\mathrm{EGRS}}}

\newcommand{\vvv}{{{\mathbf{v}}}}

\newcommand{\supp}{{\mathrm{supp}}}
\newcommand{\wt}{{{\rm{wt}}}}

\makeatletter

\newcommand{\Rmnum}[1]{\expandafter\@slowromancap\romannumeral #1@}
\newcommand\dsb[1]{\llbracket #1 \rrbracket}
\newcommand\Dsb[1]{\left\llbracket #1 \right\rrbracket}
\makeatother

\usepackage[colorlinks, linkcolor=red, citecolor=blue]{hyperref}
\usepackage{amsmath}
\usepackage{amssymb}
\usepackage{bm}
\usepackage{amssymb}
\usepackage{amsthm}
\usepackage{multirow,booktabs}
\usepackage{cases}
\usepackage{tabularx}
\usepackage{adjustbox}
\usepackage{multirow}
\usepackage[figuresright]{rotating}
\usepackage{longtable}
\usepackage{booktabs}
\usepackage{multirow}
\usepackage{color}
\usepackage{cite}
\usepackage{multirow,booktabs}
\usepackage{cases}
\usepackage{tabularx}
\usepackage{adjustbox}
\usepackage{makecell}
\usepackage{stmaryrd}
\usepackage[figuresright]{rotating}
\usepackage{longtable}
\usepackage{booktabs}
\usepackage{multirow}
\usepackage{color}
\usepackage{lineno,hyperref,mathtools}
\usepackage{soul}
\usepackage{threeparttable}
\usepackage{booktabs}
\usepackage{pifont}
\usepackage{pgfplots}
\usepackage{amsmath}
\usepackage{float}
\usepackage{bm}

\pgfplotsset{compat=1.18}
\allowdisplaybreaks   
\theoremstyle{definition} 

\newtheorem{theorem}{Theorem}

\newtheorem{problem}[theorem]{Problem}
\newtheorem{lemma}[theorem]{Lemma}
\newtheorem{corollary}[theorem]{Corollary}

\newtheorem{remark}[theorem]{Remark}
\newtheorem{example}[theorem]{Example}
\newtheorem{definition}[theorem]{Definition}

\newcommand{\edit}[1]{{\color{red}#1}}

\begin{document}
\begin{sloppypar}

\title{Plotkin-optimal pure quantum LRCs with large minimum distances from NMDS codes 
}
\author{Yang Li, Shitao Li, Gaojun Luo, San Ling
\thanks{
    This research of Yang Li and San Ling was supported by Nanyang Technological University under Research Grant 04INS000047C230GRT01. 
    The research of Shitao Li was supported by the National Natural Science Foundation of China under Grant 12526612. 
    The research of Gaojun Luo was supported by the National Natural Science Foundation of China under Grant 12401690. 
    {\em (Corresponding author: Gaojun Luo)}
} 
\thanks{Yang Li is with the School of Physical and Mathematical Sciences, Nanyang Technological University, Singapore 637371, Singapore (e-mail: yanglimath@163.com).}
\thanks{Shitao Li is with the School of Internet, Anhui University, Hefei, Anhui 230039, China (e-mail: lishitao0216@163.com).}
\thanks{Gaojun Luo is with the School of Mathematics, Nanjing University of Aeronautics and Astronautics, Nanjing, Jiangsu 211106, China (e-mail: gaojun\_luo@nuaa.edu.cn)}
\thanks{San Ling is with the School of Physical and Mathematical Sciences, Nanyang Technological University, 21 Nanyang Link, Singapore 637371, Singapore (e-mail: lingsan@ntu.edu.sg). 
He is also with VinUniversity, Vinhomes Ocean Park, Gia Lam, Hanoi 100000, Vietnam (e-mail: ling.s@vinuni.edu.vn).}
}

\maketitle

\begin{abstract}
Quantum locally recoverable codes (qLRCs) are a class of quantum error-correcting codes 
that can recover an erased qudit by accessing only a few other qudits. 
This paper develops a general framework for constructing optimal pure qLRCs that attain the Plotkin-like bound, 
referred to as Plotkin-optimal pure qLRCs, from Hermitian dual-containing near maximum-distance separable (NMDS) codes. 
Focusing on a special class of NMDS codes called extended subcodes of generalized Reed-Solomon (ESGRS) codes, 
we determine their localities by analyzing certain subset sum problems over finite fields. 
By further exploiting the algebraic structures of ESGRS codes and related codes, we construct four families of 
Plotkin-optimal pure qLRCs with large lengths and more flexible minimum distances than existing constructions. 
These constructions provide strong global error-correction capability while retaining local recovery. 
To the best of our knowledge, they give the first families of qLRCs 
attaining the pure Plotkin-like bound.  
In addition, by using the resulting locality property, we resolve in 
the negative a related problem on ESGRS codes supporting combinatorial designs. 
\end{abstract}


\section{Introduction}\label{sec.introduction}

Let $\F_q$ denote the {\em finite field} of size $q$ and let $\F_q^*=\F_q\setminus \{0\}$ denote its {\em multiplicative group}, 
where $q=p^h$ is a prime power. 
An $[n,k,d]_q$ {\em linear code} $\C$ is a $k$-dimensional linear subspace of $\F_q^n$ with minimum distance $d=d(\C)$. 
For any $\xxx=(x_1,x_2,\ldots,x_n)$ and $\yyy=(y_1,y_2,\ldots,y_n)\in \F_q^n$ (resp. $\F_{q^2}^n$), 
we define their {\em (Euclidean) inner product} (resp. {\em Hermitian inner product}) as 
$$\langle \xxx,\yyy \rangle = \sum_{i=1}^n x_iy_i~\left({\rm resp.}~
\langle \xxx,\yyy \rangle_{\rm H} = \sum_{i=1}^n x_iy_i^q\right),$$ respectively.    
Then given an $[n,k]_q$ (resp. $[n,k]_{q^2}$) linear code $\C$, 
its {\em (Euclidean) dual} (resp. {\em Hermitian dual}) is given by 
\begin{align*}
  \begin{split}
     \C^{\perp}=\left\{\yyy\in \F_q^n:~\langle \xxx,\yyy \rangle=0,~\forall~ \xxx\in \C\right\}~ 
 \left({\rm resp.} \C^{\perp_{\rm H}}=\left\{\yyy\in \F_{q^2}^n:~\langle \xxx,\yyy \rangle_{\rm H}=0,~\forall~ \xxx\in \C\right\}\right).      
  \end{split}
\end{align*}
If $d(\C)=n-k$ and $d(\C^{\perp})=k$, we say that 
$\C$ is a {\em near maximum distance separable (NMDS) code}. 
In addition, $\C$ is said to be {\em self-orthogonal} if $\C\subseteq \C^{\perp}$,  {\em dual-containing} if $\C^{\perp}\subseteq \C$,   
and these two definitions also apply when the Hermitian dual is used in place of the Euclidean dual. 
Furthermore, $\C$ is Euclidean (resp. Hermitian) self-orthogonal if and only if 
$\C^{\perp}$ (resp. $\C^{\perp_{\rm H}}$) is Euclidean (resp. Hermitian) dual-containing.

\subsection{Quantum locally recoverable codes}

Both self-orthogonal codes and dual-containing codes are closely related to quantum error-correcting codes \cite{KKKS2006}.
Let $\mathbb{C}$ be the {\em field of complex numbers} and let $\mathbb{C}^q$ be the $q$-dimensional {\em Hilbert space} over $\mathbb{C}$. 
Specifically, a {\em quantum error-correcting code} $\mathcal{Q}$ of length $n$, dimension $\kappa$, minimum distance $\delta$, and alphabet size $q$, 
denoted by $\dsb{n,\kappa,\delta}_q$, is a $q^{\kappa}$-dimensional subspace of the Hilbert space 
$(\mathbb{C}^q)^{\otimes n}=\mathbb{C}^q\otimes\mathbb{C}^q\otimes\cdots\otimes \mathbb{C}^q$. 
Such a quantum error-correcting code $\mathcal{Q}$ can encode $\kappa$ logical qudits into entangled states of $n$ qudits 
and protects against the erasure of any set of $\delta - 1$ qudits. 
Moreover, in the literature of quantum error control, $\mathcal{Q}$ is {\em pure} if distinct elements of the corresponding group 
of quantum error operators produce orthogonal results when applied to the codewords of $\mathcal{Q}$, and {\em impure} otherwise.
Similar to classical locally recoverable codes (cLRCs) \cite{GHSY2012}, Golowich and Guruswami \cite{GG-QLRC2025} 
introduced quantum locally recoverable codes (qLRCs), motivated by their potential applications in 
practically achievable large-scale quantum data storage systems \cite{B2017,J2024}. 
Generally speaking, a {\em qLRC} is a quantum error-correcting code $\mathcal{Q}$ with the property that, 
if any single qudit of $|\varphi\rangle \in \mathcal{Q}$ is erased, 
the state $|\varphi\rangle$ can be recovered from $r$ other qudits via a recovery channel. 

In the same paper \cite{GG-QLRC2025}, Golowich and Guruswami further investigated the implications of qLRCs for 
the construction and analysis of quantum low-density parity-check (qLDPC) codes. 
In a quantum locally recoverable process, each codeword is checked by only two low-weight stabilizers, 
whereas in a qLDPC code, each codeword must be constrained by a large collection of low-weight stabilizers. 
From this perspective, qLRCs may be viewed as a foundational step toward understanding and developing stronger locality properties in qLDPC codes. 
Recently, qLRCs were further generalized to the concept of $(r,\rho)$-qLRCs in \cite{GHMM2024} and they are the same if $\rho=2$. 
Note also that the connection between qLRCs and cLRCs has been established in \cite{GG-QLRC2025,GHMM2024} 
under the Euclidean and Hermitian inner products, via the CSS construction and the Hermitian construction, respectively. 
In this paper, {\em we restrict our attention to standard qLRCs within the framework of~\cite{GG-QLRC2025}}, 
as they are more directly relevant to practical applications.
For more constructions and variations of qLRCs, we refer the reader to 
\cite{BGL2025,CZ2025-TIT,CZ2026,ZC2025,GHMM2026-impure,GHMR2025,GG-QLRC2025,GHMM2024,LCEL2025,LLLL2025,LLLLL2025,XZS2025,SRT2025}.

\subsection{Known results and our motivations}

Our motivations come from two problems concerning the construction of qLRCs. 
To evaluate the performance of an $\dsb{n,\kappa,\delta}_q$ qLRC with locality $r$, 
Golowich and Guruswami~\cite{GG-QLRC2025} used quantum mechanical arguments to establish a Singleton-like bound. 
Subsequently, based on the connections between qLRCs and cLRCs, Galindo $et~al.$ \cite{GHMM2024} and Li $et~al.$\cite{LLLL2025} 
established several bounds for pure qLRCs constructed from the Hermitian construction, 
including the pure Singleton-like bound in \eqref{eq.pure_Singleton}, 
the pure Griesmer-like bound in \eqref{eq.pure_Griesmer}, and the pure Plotkin-like bound in \eqref{eq.pure_Plotkin} 
(see Lemma \ref{lem.pure_bounds} for more details). 
Furthermore, Li $et~al.$ \cite{LLLL2025} showed that such pure bounds are always tighter than 
Golowich and Guruswami's Singleton-like bound. 
Given any bound $\texttt{\textbf{B}}$ for qLRCs and any two parameters chosen from $n$, $k$, and $\delta$, 
if the remaining parameter attains the largest integer satisfying $\texttt{\textbf{B}}$ (when $\texttt{\textbf{B}}$ is an upper bound) 
or the smallest integer satisfying $\texttt{\textbf{B}}$ (when $\texttt{\textbf{B}}$ is a lower bound), 
then the corresponding qLRC is said to be {\em optimal} with respect to $\texttt{\textbf{B}}$. 
In particular, a qLRC whose parameters achieve equality in any bound for qLRCs is considered optimal.
Using Hermitian dual-containing MDS codes, Hamming codes, generalized Reed-Muller codes, projective Solomon-Stiffler codes, 
matrix-product codes, cyclic codes, trace codes, and decomposition of parity-check matrices,  
the authors in \cite{GHMM2024,LLLL2025,LCEL2025,XZS2025,CZ2025-TIT,CZ2026,ZC2025} constructed several families of pure qLRCs 
that are optimal with respect to the pure Sphere-packing-like bound, the pure Singleton-like bound, 
or the pure Griesmer-like bound. 
However, as summarized in \cite[Section \Rmnum{5}]{LLLL2025}, 
it remains open how to construct optimal pure qLRCs from the Hermitian construction 
with respect to the pure Plotkin-like bound, namely, 
\emph{Plotkin-optimal pure qLRCs}. 
This question was further proposed by Cao and Zhou in the language of matrix-product codes 
as a problem in \cite[Section \Rmnum{7}]{CZ2025-TIT}. 
Since linear codes can be viewed as matrix-product codes in a broad sense, 
we rephrase this problem in the following form.

\begin{problem}{\rm (\!\! \cite{CZ2025-TIT,LLLL2025})}\label{prob.1}
    Can we construct Plotkin-optimal pure qLRCs from the Hermitian construction?
\end{problem}

On the other hand, we observe that the known qLRCs with relatively small lengths constructed in 
\cite{GG-QLRC2025,SRT2025} via the CSS construction could have large minimum distances, 
but they are generally not optimal with respect to the corresponding bounds for qLRCs 
\cite{LCEL2025,GG-QLRC2025}. 
Furthermore, the known families of optimal pure qLRCs $\dsb{n,\kappa,\delta}_q$ with large lengths no less than $\Omega(q^2)$ 
reported in the literature typically have fixed and small minimum distances. 
For example, the largest minimum distance of the optimal pure qLRCs constructed in 
\cite{LCEL2025} is $6$, while those of the optimal pure qLRCs constructed in 
\cite{LLLL2025,XZS2025,CZ2025-TIT,CZ2026,ZC2025} are only $3$ or $4$. 
As far as we know, only the family of optimal pure qLRCs contained in \cite[Theorem \edit{9}]{LLLLL2025}
has both large lengths and minimum distances. 
However, this family is defined only over finite fields of characteristic $2$, and its dimensions are restricted to even values. 
We summarize the specific parameters of these qLRCs with large lengths in Table~\ref{tab:qLRCs} for easy reference. 
Note also that, for better correction of qudit erasures and errors, 
it is desirable to construct {\em qLRCs with larger and more flexible minimum distances}, 
as {such properties provide greater freedom in balancing the parameters required for both local and global quantum recovery \cite{LLLLL2025}}.  
In view of the above observations, the following problem naturally arises.

\begin{problem}\label{prob.2}
    Can we construct optimal pure qLRCs with large lengths and flexible minimum distances over general finite fields via the Hermitian construction?
\end{problem}

\subsection{Our contributions and techniques}
This paper addresses {\bf Problem} \ref{prob.1} and {\bf Problem} \ref{prob.2} in a unified manner 
by exploiting a special family of NMDS codes, namely the extended subcodes of generalized Reed-Solomon (ESGRS) codes 
$\ESGRS_k(\SSS,\mathbf{v},\infty)$ introduced in \cite{ADV2025} and \cite{LSZ2025}, 
whose formal definition can be found in Definition~\ref{def.ESGRS_code}.
In particular, we leverage ESGRS codes to construct several families of Plotkin-optimal pure qLRCs 
with large lengths and flexible minimum distances. 
The main contributions and techniques of this work can be summarized as follows.
\begin{enumerate}
    \item In Theorem~\ref{th.pure_qLRCs_from_NMDS}, we show that 
    applying the Hermitian construction in Lemma~\ref{lem.Hermitian_qLRC}
    to any $[n,k,n-k]_{q^2}$ NMDS code with $2\le k\le n-2$ 
    necessarily yields a pure $\dsb{n,2k-n,n-k}_q$ qLRC $\mathcal{Q}$ with locality $r\in \{k-1,k\}$. 
    Moreover, when $r=k-1$, we prove that $\mathcal{Q}$ is simultaneously optimal with respect to
    the pure Singleton-like bound in \eqref{eq.pure_Singleton},
    the pure Griesmer-like bound in \eqref{eq.pure_Griesmer},
    and the pure Plotkin-like bound in \eqref{eq.pure_Plotkin}.
    In contrast, when $r=k$, we establish additional conditions ensuring the optimality of $\mathcal{Q}$. 
    As stated in Remark~\ref{rem.lmds}, the purity properties can also be extended to so-called $\ell$-MDS codes.
    In particular, Theorem~\ref{th.pure_qLRCs_from_NMDS} provides a general framework toward addressing {\bf Problem}~\ref{prob.1}.

    \item For any $[n+1,k,n-k+1]_{q}$ ESGRS code $\ESGRS_k(\SSS,\mathbf{v},\infty)$, 
    we first show in Theorem~\ref{th.ESGRS_cLRCs}.1) that it has locality $k$ by 
    explicitly identifying singular submatrices in its generator matrix.  
    By considering solutions to certain subset sum problems over finite fields, 
    we then obtain in Theorem~\ref{th.ESGRS_cLRCs}.2) a sufficient condition under which 
    the dual code $\ESGRS_k(\SSS,\mathbf{v},\infty)^{\perp}$ has locality $n-k$ in Theorem \ref{th.ESGRS_cLRCs}.2). 
    Although it is trivial for a $k$-dimensional linear code to have locality $k$, 
    we exploit this locality property of ESGRS codes to further show that the minimum weight codewords of 
    $\ESGRS_k(\SSS,\mathbf{v},\infty)$ cannot support nontrivial combinatorial $t$-designs for any $t\geq 1$, 
    thereby resolving {\em in the negative} a problem in \cite{LSZ2025}. 
    Such a result is established in Theorem~\ref{th.designs_locality}.

    \item By further analyzing the algebraic structures of ESGRS codes together with subset sum problems over finite fields, 
    we apply the framework developed in Theorem~\ref{th.pure_qLRCs_from_NMDS} to construct three families of Plotkin-optimal pure qLRCs 
    in Theorems~\ref{th.qlrc111}, \ref{th.qlrc222}, and~\ref{th.qlrc333}. 
    Another family of Plotkin-optimal pure qLRCs is obtained in Remark \ref{rem.second-family}.2) as a complement of Theorem \ref{th.qlrc111}. 
    Note that these qLRCs also achieve the pure Singleton-like bound in~\eqref{eq.pure_Singleton} and  
    the pure Griesmer-like bound in~\eqref{eq.pure_Griesmer}. 
    For easy reference, we collect their parameters in Table \ref{tab:our_qLRCs}. 
    Relative to the known results listed in Table~\ref{tab:qLRCs},
    our optimal qLRCs admit not only lengths of order $\Omega(q^2)$,
    but also larger and more flexible minimum distances. 
    Therefore, Theorems~\ref{th.qlrc111}, \ref{th.qlrc222}, and \ref{th.qlrc333}, as well as Remark \ref{rem.second-family} provide affirmative answers to
    {\bf Problem}~\ref{prob.1} and {\bf Problem} \ref{prob.2} at the same time.
\end{enumerate}

\begin{table*}
\centering
\caption{Known families of optimal pure qLRCs with large lengths}
\label{tab:qLRCs}
\setlength{\tabcolsep}{3pt}
\renewcommand{\arraystretch}{1.3}

\resizebox{\textwidth}{!}{
\begin{tabular}{c|c|c|c|c|l}
\toprule
No. & $\dsb{n,\kappa,\delta}_q$ 
& Maximum length 
& $r$ 
& Condition 
& Reference\\ 
\midrule

1 & 
$\Dsb{Nt(2q+v-2), Nt(2q+v-2)-4Nt-2, \edit{3}}_{q}$ 
& $\Omega(q^2)$
& $q+v-2$
& \makecell[c]{$N\mid(q^2-1)$, $N\nmid(q+1)$, \\ 
$q\geq 8$, $v\mid(q-1)$, \\
$Nvt\leq q+v-4\leq q$}
& \cite[Theorem 9]{CZ2025-TIT}
\\ \hline

2 & 
$\Dsb{n=(2m+1)(2q+v-2)t,n-4(2m+1)t-2,\edit{3}}_{q}$ 
& $\Omega(q^2)$
& $q+v-2$
& \makecell[c]{$2m\mid(q^2-1)$, odd $q\geq 9$, \\ 
$v\mid(q-1)$, $v\geq 6$, \\
$(2m+1)vt\leq q+v-4$}
& \cite[Theorem 29]{CZ2026}
\\ \hline 

3 & 
$\Dsb{mh,mh-2m-4,\edit{4}}_{q}$ 
& $\Omega(q^2)$ 
& $h-1$
& \makecell[c]{$m,h\leq q$, $(h-1)\mid(q-1)$,\\
$m\geq 4$, $h\geq 5$, odd $q$, \\
$1-h$ is a square in $\mathbb F_q$}
& \cite[Table~I]{GHMR2025}
\\ \hline


4 & 
$\Dsb{n=\frac{q^{2m}-1}{q^2-1}-\sum_{i=1}^s\frac{q^{2u_i}-1}{q^2-1}, n-2m, \edit{3}}_q$
& $\Omega(q^{2m-2})$
& $q^{2m-2}-\sum_{i=1}^sq^{2u_i-2}-1$
& \makecell[c]{$u_i$, $m$, and $q$ satisfy \\ some specific conditions}
& \cite[Theorem 10]{LLLL2025} 
\\ \hline 

5 & 
$\dsb{u(r+1),u(r-1)-2\delta+2,\edit{\delta}}_{q}$ 
& $\Omega(q^2)$ 
& \makecell[c]{$2\delta+u-4<r\leq q-1$} 
& \makecell[c]{$\edit{2}\leq \delta\leq \edit{6}$} 
& \cite[Theorem 11]{LCEL2025}  
\\ \hline 

6 & 
$\dsb{q^2,q^2-6,\edit{3}}_{q}$ 
& $\Omega(q^2)$  
& $q^2-q-1$  
& $q=p^h\geq 7$ 
& \cite[Theorem 6]{XZS2025} 
\\ \hline 

7 & 
$\Dsb{t(2q+v-2), t(2q+v-2)-4t-4, \edit{4}}_{q}$ 
& $\Omega(q^2)$
& $q+v-2$
& \makecell[c]{$q\geq 9$, $v\mid(q-1)$,\\
$v\geq 8$, $vt\leq q+v-5$}
& \cite[Theorem 5.6]{ZC2025}
\\

\bottomrule
\end{tabular}
}

\vspace{1mm}
\begin{minipage}{0.96\textwidth}
\footnotesize
\raggedright
\textit{Note.}
The authors constructed more families of optimal pure qLRCs in 
\cite{CZ2025-TIT,CZ2026,ZC2025,GHMR2025,LLLL2025}. 
For each reference, we list one representative family with the largest minimum distance. 
The reader is referred to the corresponding references for more details on the other constructions.
\end{minipage}

\end{table*}

\begin{table*}
\centering
\caption{Our Plotkin-optimal pure qLRCs with large lengths and flexible minimum distances}
\label{tab:our_qLRCs}
\setlength{\tabcolsep}{3pt}
\renewcommand{\arraystretch}{1.3}

\begin{tabular}{c|c|c|c|c|c}
\toprule
No. & $\dsb{n,\kappa,\delta}_q$ 
& Maximum length  
& $r$ 
& Condition 
& Reference\\ 
\midrule

1 & 
$\Dsb{q^2,q^2-2k,\edit{k}}_q$ 
& $\Omega(q^{2})$  
& $q^{2}-k-1$ 
& \makecell[c]{$q=p^h\geq 5$, $p\geq 3$, 
$\edit{\left\lceil \frac{2p}{p-2} \right\rceil}\leq k\leq \edit{q-2}$}   
& Theorem~\ref{th.qlrc111} 
\\ \hline

2 & 
$\Dsb{4^m,4^m-2k,\edit{k}}_{2^m}$ 
& $\Omega(4^m)$  
& $4^m-k-1$ 
& $\edit{2\mid k}$, $\edit{4}\leq k\leq \edit{2^m-2}$, $m\geq 3$ 
& Remark \ref{rem.second-family}  
\\ \hline

3 & 
$\Dsb{q^2-\ell,q^2-2k-\ell,\edit{k}}_q$ 
& $\Omega(q^{2})$  
& $q^{2}-k-\ell-1$ 
& \makecell[c]{$q=p^h\geq 7$, $1\leq \ell\leq q-5$, 
$\edit{3}\leq k\leq \edit{q-\ell-2}$,\\
$\ell \leq 2k-(p+1)\left\lfloor \frac{k}{p} \right\rfloor
+\log_q \left(\frac{p^{\left\lfloor \frac{k}{p} \right\rfloor}}{2^k}\right)-[k]_p-2$}        
& Theorem~\ref{th.qlrc222} 
\\ \hline

4 & 
$\Dsb{q^2+1,q^2-2k+1,\edit{k}}_{q}$ 
& $\Omega(q^2)$  
& $q^2-k$  
& \makecell[c]{$q=p^h\geq 4$, 
$\edit{3}\leq k\leq \edit{q-1}$, $k\neq q-2$} 
& Theorem~\ref{th.qlrc333} 
\\ 

\bottomrule
\end{tabular}

\vspace{1mm}
\begin{minipage}{0.96\textwidth}
\footnotesize
\raggedright
\textit{Note.}
These four families of Plotkin-optimal pure qLRCs are also optimal with respect to the pure Singleton-like bound 
in \eqref{eq.pure_Singleton} and the pure Griesmer-like bound in \eqref{eq.pure_Griesmer}, simultaneously. 
The notation used in the table is specified in the corresponding theorems.
\end{minipage}

\end{table*}

This paper is organized as follows.
Section~\ref{sec.2} reviews several known results on qLRCs. 
Section~\ref{sec.3} establishes a general framework for constructing Plotkin-optimal pure qLRCs from Hermitian dual-containing NMDS codes.
Section~\ref{sec.4} gives a brief introduction to ESGRS codes and investigates their locality properties.
Leveraging these locality properties, we further resolve in the negative a problem concerning combinatorial designs supported by ESGRS codes.
Section~\ref{sec.5} subsequently derives four families of Plotkin-optimal pure qLRCs with large lengths and flexible minimum distances 
via ESGRS codes.
Finally, Section~\ref{sec.6} concludes the paper with some remarks and directions for future research.

\section{Known results on qLRCs}\label{sec.2}

We first recall some basic results on cLRCs. 
Given a codeword $\ccc=(c_1,c_2,\ldots,c_n)\in \C$, 
define its {\em support} as $\supp(\ccc)=\{1\leq i\leq n:~c_i\neq 0\}$. 
Then a cLRC can be mathematically defined as follows. 

\begin{definition}{\rm (\!\! \cite[Definition 1]{LCEL2025})}
  An $[n,k,d]_q$ linear code $\C$ with dual $\C^{\perp}$ is said to be a 
  {\em classical locally recoverable code (cLRC) with locality $r$} 
  if, for each $i\in \{1,2,\ldots,n\}$, 
  there exists some codeword $\overline{\ccc}=(c_1,c_2,\ldots,c_n)\in \C^{\perp}$ 
  such that $i\in \supp(\overline{\ccc})$ and $|\supp(\overline{\ccc})|\leq r+1$. 
  Moreover, we abbreviate such a code $\C$ as an {$(n,k,d,q;r)$ cLRC}. 
\end{definition}

In fact, for a given NMDS code, its locality can only take two possible values, as shown in the following lemma. 

\begin{lemma}{\rm (\!\! \cite[Theorem 14]{TFDTZ2023})}\label{lem.NMDS_cLRC}
    Let $\C$ be an $[n,k,n-k]_q$ NMDS code with $2 \leq k \leq n-2$. 
    Then the locality of $\C$ is either $k-1$ or $k$.
\end{lemma}


Similar to cLRCs, qLRCs are quantum error-correcting codes with a specific locality property. 
Note that there are various types of quantum error-correcting codes, and the so-called {\em quantum stabilizer codes} 
form the most important and interesting class. 
Quantum stabilizer codes have close connections to classical codes, 
containing those derived from the CSS construction, Hermitian construction, and symplectic construction \cite{KKKS2006}. 
Compared with the CSS construction, the Hermitian construction typically produces quantum error-correcting codes with superior parameters, 
and relative to the symplectic construction, it is generally more convenient to apply. 

In this paper, we mainly focus on qLRCs obtained from the Hermitian construction.  
Then it is more straightforward to consider the definition of qLRCs under the notation used in \cite{GHMM2024}. 

\begin{definition}{\rm (\!\! \cite[Definitions 9 and 10]{GHMM2024})}\label{def.qLRCs}
  A quantum error-correcting code $\mathcal{Q}\subseteq (\mathbb{C}^q)^{\otimes n}$ is said to be a 
  {\em quantum locally recoverable code (qLRC) with locality $r$} if, 
  for each $i\in \{1,2,\ldots,n\}$, there exists a set $J\subseteq \{1,2,\ldots,n\}$ containing $i$ with $|J|\leq r+1$ 
  such that for each subset $I\subsetneq J$ with $|I|=1$, there exists a trace-preserving map $\mathcal{R}_{\mathcal{Q},I}^J$, 
  which acts only on the qudits corresponding to $J$ and keeps untouched the remaining ones, such that 
  $$
  \mathcal{R}_{\mathcal{Q},I}^J \circ \Gamma^I(|\varphi\rangle \langle \varphi|) = |\varphi\rangle \langle \varphi|
  $$
  for any $|\varphi\rangle \in \mathcal{Q}$, where $\Gamma^I$ is a mapping given as in \cite{GHMM2024}.
\end{definition}

In particular, for $\mathcal{Q}$ being a stabilizer code, Definition \ref{def.qLRCs} can be reduced to an easier form 
by considering cLRCs with special structures. 
Let $\wt(\SSS)$ denote the minimum weight of all nonzero vectors in any nonempty set $\SSS\subseteq \F_{q^2}^n$. 
The following lemma gives the Hermitian construction for qLRCs.


\begin{lemma}\label{lem.Hermitian_qLRC}
{\rm (\!\! Hermitian construction for qLRCs \cite[Theorem 29]{GHMM2024})} 
  If $\C$ is an $[n,k,d]_{q^2}$ Hermitian dual-containing code with locality $r$ and $d(\C^{\perp_{\rm H}})\geq 2$, 
then there exists an 
$$\Dsb{n,\kappa,\delta}_q$$ 
qLRC $\mathcal{Q}$ with locality $r$, 
where 
\begin{align*}
    \kappa = 2k-n, \quad \delta = \left\{
    \begin{array}{ll} 
        \wt(\C\setminus \C^{\perp_{\rm H}})\geq d, & \mbox{if}~ \C^{\perp_{\rm H}}\subsetneq \C, \\
        d, & \mbox{if}~ \C^{\perp_{\rm H}}= \C. 
    \end{array}
        \right.
\end{align*}
Furthermore, such a qLRC $\mathcal{Q}$ is said to be {\em pure} if $\delta = d$; and {\em impure} otherwise.
\end{lemma}

\begin{lemma}{\rm (\!\! \cite[Theorem 31]{GHMM2024} and \cite[Theorem \edit{4}]{LLLL2025})}\label{lem.pure_bounds}
    Let $\mathcal{Q}$ be a pure $\dsb{n,\kappa,\delta}_q$ qLRC with locality $r$ constructed by the Hermitian construction 
    given in Lemma \ref{lem.Hermitian_qLRC}. 
    Let $\Delta=\left\lceil \frac{n+\kappa}{2r} \right\rceil-1$. 
    Then the following bounds hold. 
    \begin{enumerate}
        \item {\em (Pure Singleton-like bound)} We have that 
        \begin{align}\label{eq.pure_Singleton}
            2\delta \leq n - \kappa - 2\Delta + 2.
        \end{align}

        \item {\em (Pure Griesmer-like bound)}
        We have that  
        \begin{align}\label{eq.pure_Griesmer}
            n \geq \max_{0\leq \ell\leq \Delta}
                \left\{\ell(r+1)+\sum_{t=0,~2\mid t}^{n+\kappa-2\ell r-2} \left\lceil \frac{\delta}{q^{t}} \right\rceil \right\}.
        \end{align}

        \item {\em (Pure Plotkin-like bound)}
        We have that 
        \begin{align}\label{eq.pure_Plotkin}
            \delta \leq \min_{0\leq \ell\leq \Delta}
                \left\{ \frac{q^{n+\kappa - 2\ell r - 2}(q^2 - 1)(n - \ell(r+1))}{q^{n+\kappa - 2\ell r} - 1} \right\}.
        \end{align}
    \end{enumerate}
\end{lemma}

\begin{remark}\label{rem.CSS_bounds}
Although the bounds presented in Lemma~\ref{lem.pure_bounds} were originally derived for the Hermitian construction, 
it is straightforward to verify that they remain valid for pure qLRCs obtained from the CSS construction (see \cite[Corollary~3]{LCEL2025}) as well. 
In this case, it suffices to replace $q$ by $q^2$ in the corresponding formulas of these bounds, as linear codes over $\F_{q^2}$ are considered here.
\end{remark}

\section{A general framework to construct Plotkin-optimal pure qLRCs from NMDS codes}\label{sec.3}

This section develops a general framework for resolving {\bf Problem} \ref{prob.1} by employing Hermitian dual-containing NMDS codes. 
Within this framework, we establish the purity of the resulting qLRCs and analyze their optimality.

\begin{theorem}\label{th.pure_qLRCs_from_NMDS}
    Let $\C$ be an $[n,k,n-k]_{q^2}$ Hermitian dual-containing NMDS code with $2 \leq k \leq n-2$. 
    Then the following statements hold. 
    \begin{enumerate}
        \item Applying the Hermitian construction in Lemma \ref{lem.Hermitian_qLRC} to $\C$ yields 
        a pure $\dsb{n,2k-n,n-k}_q$ qLRC $\mathcal{Q}$ with locality 
        $r\in \{k-1,k\}$. 


        \item If $\mathcal{Q}$ has locality $k-1$, then $\mathcal{Q}$ is optimal with respect to 
        the pure Singleton-like bound in \eqref{eq.pure_Singleton}, 
        the pure Griesmer-like bound in \eqref{eq.pure_Griesmer}, 
        and the pure Plotkin-like bound in \eqref{eq.pure_Plotkin}, simultaneously.
        
        \item If $\mathcal{Q}$ has locality $k$, then $\mathcal{Q}$ is optimal 
        with respect to the pure Griesmer-like bound in \eqref{eq.pure_Griesmer} if $n>q^2+k$ 
        and the pure Plotkin-like bound in \eqref{eq.pure_Plotkin} if $0\leq k-\frac{q^{2k-2}-1}{q^{2k}-1}\cdot n< 1$.  
    \end{enumerate}
\end{theorem}
\begin{IEEEproof}
    1) Since $\C$ is an $[n,k,n-k]_{q^2}$ Hermitian dual-containing NMDS code, 
    we have $k \ge \frac{n}{2}$, and its Hermitian dual $\C^{\perp_{\rm H}}$ is an $[n,n-k,k]_{q^2}$ NMDS code.
    Applying Lemma~\ref{lem.Hermitian_qLRC} to $\C$ and then using Lemma~\ref{lem.NMDS_cLRC} yields an 
    $\dsb{n,2k-n,\delta}_q$ qLRC $\mathcal{Q}$ with locality $r \in \{k-1, k\}$, where
\begin{align*}
    \delta=\left\{
    \begin{array}{ll} 
        \wt(\C\setminus \C^{\perp_{\rm H}})\geq d(\C), & \mbox{if}~ \C^{\perp_{\rm H}}\subsetneq \C, \\
        d(\C), & \mbox{if}~ \C^{\perp_{\rm H}}= \C. 
    \end{array}    
    \right.
\end{align*}
It suffices to consider the case $\C^{\perp_{\rm H}} \subsetneq \C$, which implies $k > \frac{n}{2}$ and $d(\C^{\perp_{\rm H}}) > d(\C)$.
Hence, there exists a codeword $\mathbf{c} \in \C$ such that $\wt(\mathbf{c}) = d(\C)$ and $\mathbf{c} \notin \C^{\perp_{\rm H}}$.
As a result, we have $\ccc\in \C\setminus \C^{\perp_{\rm H}}$ and $\wt(\C\setminus \C^{\perp_{\rm H}})=d(\C)$. 
Since $d(\C)=n-k$, $\mathcal{Q}$ is a pure $\dsb{n,2k-n,n-k}_q$ qLRC with locality $r \in \{k-1, k\}$. 

    2) Since $r=k-1$ and $2\leq k\leq n-2$, we have 
    $
    \left\lceil \frac{n+\kappa}{2r} \right\rceil = \left\lceil \frac{2k}{2k-2} \right\rceil = 2. 
    $
    Then $\ell=0$ or $1$. Taking $\ell=1$ and applying the pure Singleton-like bound in \eqref{eq.pure_Singleton}, 
    the pure Griesmer-like bound in \eqref{eq.pure_Griesmer}, and the pure Plotkin-like bound in \eqref{eq.pure_Plotkin} to
    $\mathcal{Q}$ respectively gives that 
    \begin{align}\label{eq.bound1}
        \begin{split}
                        n-\kappa-2\left\lceil \frac{n+\kappa}{2r} \right\rceil+4  
         =   n-(2k-n)-4+4 
         =   2(n-k), 
        \end{split}
    \end{align}
    \begin{align}\label{eq.bound2}
        \ell(r+1)+\sum_{t=0,~2\mid t}^{n+\kappa-2\ell r-2} \left\lceil \frac{\delta}{q^{t}} \right\rceil = k+ \left\lceil \frac{n-k}{q^0} \right\rceil=n,
    \end{align}
    and 
    \begin{align}\label{eq.bound3}
        \begin{split}
                    \frac{q^{n+\kappa - 2\ell r - 2}(q^2 - 1)(n - \ell(r+1))}{q^{n+\kappa - 2\ell r} - 1} 
      =  \frac{q^0(q^2 - 1)(n - k)}{q^2 - 1} 
      =  n-k. 
        \end{split}
    \end{align}
    Comparing \eqref{eq.bound1}, \eqref{eq.bound2}, and \eqref{eq.bound3} with the parameters of $\mathcal{Q}$, 
    we conclude that $\mathcal{Q}$ is optimal with respect to these three bounds.

    3) Since $\C$ is an $[n,k,d]_{q^2}$ Hermitian dual-containing NMDS code with $2\leq k\leq n-2$, 
    it follows from \cite[Theorem 2.7]{DL2000} that $n\leq 2q^2+k$. Then $n-k\leq 2q^2\leq q^4$, as $q\geq 2$. 
    Note that $\left\lceil \frac{n+\kappa}{2r} \right\rceil=1$ and $\ell=0$. 
    Applying the pure Griesmer-like bound in \eqref{eq.pure_Griesmer} 
    and the pure Plotkin-like bound in \eqref{eq.pure_Plotkin} to $\mathcal{Q}$ gives that 
    \begin{align}\label{eq.bound4}
        \begin{split}
            n\geq       \sum_{t=0,~2\mid t}^{n+\kappa-2\ell r-2}  \left\lceil \frac{n-k}{q^t} \right\rceil
        = \sum_{t=0}^{k-1}\left\lceil \frac{n-k}{q^{2t}} \right\rceil  
         = \left\{\begin{array}{ll}
            n-1, & \mbox{if}~ n\leq q^2+k, \\
            n, & \mbox{if}~ q^2+k<n\leq 2q^2+k,
        \end{array}\right. 
        \end{split}
    \end{align}
    and 
    \begin{align}\label{eq.bound5}
       n-k\leq \frac{q^{n+\kappa - 2}(q^2 - 1)n }{q^{n+\kappa} - 1}=\frac{q^{2k-2}(q^2-1)n}{q^{2k}-1} = n- \frac{q^{2k-2}-1}{q^{2k}-1}\cdot n,
    \end{align}
    respectively. 
    Comparing \eqref{eq.bound4} and \eqref{eq.bound5} with the parameters of $\mathcal{Q}$, 
    we conclude that $\mathcal{Q}$ is optimal with respect to the pure Griesmer-like bound in \eqref{eq.pure_Griesmer} if $n>q^2+k$ 
    and the pure Plotkin-like bound in \eqref{eq.pure_Plotkin} if $k=\left\lceil \frac{q^{2k-2}-1}{q^{2k}-1}\cdot n\right\rceil$, 
    which is equivalent to $0\leq k-\frac{q^{2k-2}-1}{q^{2k}-1}\cdot n< 1$.  
\end{IEEEproof}

Note that Theorem~\ref{th.pure_qLRCs_from_NMDS} in fact provides a general framework for 
constructing Plotkin-optimal pure qLRCs from Hermitian dual-containing NMDS codes. 
To highlight this perspective, we state the following result as an immediate corollary  
of Theorem~\ref{th.pure_qLRCs_from_NMDS}.

\begin{corollary}
    Given an $[n,k,n-k]_{q^2}$ Hermitian dual-containing NMDS code with $2 \leq k \leq n-2$,  
    there exists a pure $\dsb{n,2k-n,n-k}_q$ qLRC $\mathcal{Q}$ with locality 
    $r\in \{k-1,k\}$. Moreover, $\mathcal{Q}$ is Plotkin-optimal 
    if one of the following conditions holds: 
    \begin{itemize}
        \item $r=k-1$;
        \item $r=k$ and $0\leq k-\frac{q^{2k-2}-1}{q^{2k}-1}\cdot n< 1$. 
    \end{itemize}
\end{corollary}

For an $[n,k,d]_{q^2}$ linear code $\C$, we define its {\em Singleton defect} $\SSS(\C) = n - k - d + 1$ \cite{D1996}. 
According to the well-known {Singleton bound} \cite{HP2003}, $\SSS(\C)\geq 0$. 
As a generalization of NMDS codes, we call $\C$ an {\em $\ell$-MDS code} 
if $\SSS(\C)=\SSS(\C^{\perp_{\rm H}})=\ell$, where $0\leq \ell \leq \min\{k,n-k\}$.

\begin{remark}\label{rem.lmds} 
We now consider Theorem~\ref{th.pure_qLRCs_from_NMDS} in the more general $\ell$-MDS setting.
We have the following additional remarks.
\begin{enumerate}
\item  By an argument analogous to that in the proof of Theorem~\ref{th.pure_qLRCs_from_NMDS}.1), any qLRC obtained from a 
Hermitian dual-containing $\ell$-MDS code must be pure.
In other words, if applying Lemma~\ref{lem.Hermitian_qLRC} to a code $\mathcal{C}$ results in an impure qLRC, 
then $\mathcal{C}$ is not $\ell$-MDS.

\item Note also that for $\ell=0$, an $\ell$-MDS code is more frequently called an MDS code. 
It is well-known that Hermitian dual-containing MDS codes always 
yield pure quantum MDS codes (see also \cite{R1999}). 
There are also many families of Hermitian dual-containing MDS codes constructed in the literature 
(see, e.g., \cite{JLLX2010,JX2014,JKW2017,GLL2021,FW2024,CLC2024,KZ2012,KZL2014} and the tables or references therein).
However, an $[n,k]_{q^2}$ MDS code must have locality $k$, 
which is trivial from the practical viewpoint according to \cite{GHSY2012,TB2014,LLLLL2025}. 
Therefore, it may {\em not} be very interesting to construct pure qLRCs from Hermitian dual-containing MDS codes. 
\end{enumerate}
\end{remark}


\section{Localities of extended subcodes of GRS codes}\label{sec.4}

\subsection{Introduction to extended subcodes of GRS codes}

The following gives the definition of extended subcodes of generalized Reed-Solomon (ESGRS) codes, 
which was first introduced by Li $et~al.$ in \cite{LSZ2025} 
and was further developed by Abdukhalikov $et~al.$ in \cite{ADV2025}. 

\begin{definition}\label{def.ESGRS_code}
    Let $\SSS=\{a_1,a_2,\ldots,a_n\}\subseteq \F_{q}$ and let $\mathbf{v}=(v_1,v_2,\ldots,v_n)\in (\F_{q}^*)^n$. 
    Let 
    \begin{align*}
        \mathbb{V}_{q}[x]_{k}=\left\{f(x)=\sum_{i=0}^{k-2} f_ix^i+ f_{k}x^{k}:~ f_0,f_1,\ldots,f_{k-2},f_k\in \F_q\right\}.
    \end{align*}
    Given $3\leq k\leq n-2\leq q-2$, an $[n+1,k]_{q}$ {\em extended subcode of generalized Reed-Solomon (ESGRS) code} is defined as  
     \begin{align*}
        \begin{split}
            \ESGRS_k(\SSS,\mathbf{v},\infty) =  \Bigl\{(v_1f(a_1),v_2f(a_2),\ldots,v_n & f(a_n), f_{k}):~f(x)\in \mathbb{V}_{q}[x]_{k}\Bigr\},  
        \end{split}
    \end{align*} 
    where $f_k$ is the coefficient of $x^k$ in $f(x)\in \mathbb{V}_{q}[x]_{k}$. 
    Moreover, it can be checked that $\ESGRS_k(\SSS,\mathbf{v},\infty)$ has a generator matrix of the form  
\begin{equation}\label{eq.ETGRS.generator_matrix}
        G_{\ESGRS_k(\SSS,\mathbf{v},\infty)}=\left(\begin{array}{ccccc}
        v_1  & v_2  & \cdots & v_n & 0\\
        v_1a_1  & v_2a_2  & \cdots & v_na_n & 0\\
        \vdots  & \vdots  & \ddots & \vdots & \vdots\\
        v_1a^{k-2}_1  & v_2a^{k-2}_2  & \cdots & v_na^{k-2}_n & 0\\
        v_1a_1^k & v_2a_2^k & \cdots & v_na_n^k & 1\\
        \end{array}\right).
\end{equation} 
\end{definition}

From \cite{LSZ2025}, there is an easy way to determine whether an ESGRS code is NMDS or not. 
Given a set $\SSS\subseteq \F_q$, an integer $k$ with $1\leq k\leq |\SSS|$, and $b\in \F_q$,  
we use $N(k,b,\SSS)$ to denote the number of $k$-subsets $\{x_1,x_2,\ldots,x_k\}\subseteq \SSS$ such that 
$$
x_1+x_2+\cdots+x_k=b.
$$ 
The problem of determining whether $N(k,b,\SSS)>0$ is usually called the \emph{subset sum problem}. 
In addition, it is natural to define that 
\begin{align}
    \label{eq.Nkbs}
    N(0,b,\SSS)=\left\{\begin{array}{ll}
        1, & \mbox{if}~ b=0, \\
        0, & \mbox{if}~ b\in \F_q^*.
    \end{array}\right.
\end{align}
Then we have the following result.

\begin{lemma}{\rm (\!\! \cite[Theorem 4]{LSZ2025})}\label{lem.NMDS}
    For given $\SSS$ and $\vvv$, $\ESGRS_k(\SSS,\mathbf{v},\infty)$ is an $[n+1,k,n-k+1]_q$ NMDS code if and only if      
    $N(k,0,\SSS)> 0$, and it is an $[n+1,k,n-k+2]_q$ MDS code if and only if      
    $N(k,0,\SSS)= 0$. 
\end{lemma}

\subsection{{Localities of ESGRS codes and their duals}}

For an $[n+1,k]_q$ linear code $\C$ and $0\leq t\leq n$, let  
\begin{align}\label{eq.Bk}
    \mathcal{B}_t\left(\C\right)=\left\{\supp(\ccc):~ \ccc\in \C,~ \wt(\ccc)=t\right\},   
\end{align}
where $\supp(\ccc)$ is the support of $\ccc\in \C$ introduced before. 
Using \eqref{eq.Bk}, we have the following ways to 
determine the localities of NMDS codes and their duals.

\begin{lemma}\label{lem.locality_NMDS}
    Let $\C$ be an $[n+1,k,n-k+1]_q$ NMDS code. 
    Then the following statements hold. 
    \begin{enumerate}
        \item {\rm (\!\! \cite[Corollary 1]{TFDTZ2023})} $\C$ has locality $k-1$ if and only if 
        $$\bigcup_{B\in \mathcal{B}_k\left(\C^{\perp}\right)} B=\{1,2,\ldots,n+1\}.$$

        \item {\rm (\!\! \cite[Theorem 16]{TFDTZ2023})} $\C^{\perp}$ has locality $n-k$ if 
        $$\bigcap_{B\in \mathcal{B}_k\left(\C^{\perp}\right)} B=\emptyset.$$
    \end{enumerate}    
\end{lemma}

\begin{theorem}\label{th.ESGRS_cLRCs}
    For given $\SSS=\{a_1,a_2,\ldots,a_{n}\}\subseteq \F_q$ and $\vvv=(v_1,v_2,\ldots,v_{n})\in (\F_q^*)^{n}$, 
    if $N(k,0,\SSS)> 0$, then the following statements hold. 
    \begin{enumerate}
        \item $\ESGRS_k(\SSS,\mathbf{v},\infty)$ is an $(n+1,k,n-k+1,q;k)$ cLRC. 
        
        \item If furthermore $\sum_{i=0}^{k}(-1)^i N(k-i,-ia_j,\SSS)> 0$ for any $a_j\in \SSS$, then 
        $\ESGRS_k(\SSS,\mathbf{v},\infty)^{\perp}$ is an $(n+1,n-k+1,k,q;n-k)$ cLRC. 
    \end{enumerate}
\end{theorem}
\begin{IEEEproof}
    It follows from Lemma \ref{lem.NMDS} that $\ESGRS_k(\SSS,\mathbf{v},\infty)$ is an $[n+1,k,n-k+1]_q$ NMDS code. 
    Let $\ccc^{\perp}\in \ESGRS_k(\SSS,\mathbf{v},\infty)^{\perp}$ be a codeword with weight $k$, and 
    let $\ccc_k^{\perp}$ be the corresponding nonzero vector of length $k$ obtained by deleting the zero coordinates of $\ccc^{\perp}$. 
    Denote by $M_{k}$ a $k\times k$ submatrix obtained by selecting $k$ distinct columns of $G_{\ESGRS_k(\SSS,\mathbf{v},\infty)}$ 
    indexed by $\{i_1,i_2,\ldots,i_k\}$, where $G_{\ESGRS_k(\SSS,\mathbf{v},\infty)}$ is given in \eqref{eq.ETGRS.generator_matrix} 
    and $1\leq i_j\leq n+1$ for any $1\leq j\leq k$.  
    Since the matrix $G_{\ESGRS_k(\SSS,\mathbf{v},\infty)}$ is a parity-check matrix of 
    $\ESGRS_k(\SSS,\mathbf{v},\infty)^{\perp}$, then 
    the minimum weight codeword $\ccc^{\perp}$ has support $\{i_1,i_2,\ldots,i_k\}$ if and only if 
    \begin{align}\label{eq.locality_condition}
        M_{k}\cdot (\ccc_k^{\perp})^T=\mathbf{0}.
    \end{align}

    According to the structure of $G_{\ESGRS_k(\SSS,\mathbf{v},\infty)}$ given in \eqref{eq.ETGRS.generator_matrix}, 
    $M_{k}$ has the following two possible forms 
    \begin{align*}
        M_{k}^\Rmnum{1}=\left(\begin{array}{ccccc}
        v_{i_1}  & v_{i_2}  &  \cdots & v_{i_{k-1}} & 0 \\
        v_{i_1}a_{i_1}  &  v_{i_2}a_{i_2}  & \cdots & v_{i_{k-1}}a_{i_{k-1}} & 0 \\
        \vdots  & \vdots  & \ddots & \vdots & \vdots \\
        v_{i_1}a^{k-2}_{i_1}  &  v_{i_2}a^{k-2}_{i_2}  & \cdots & v_{i_{k-1}}a^{k-2}_{i_{k-1}} & 0 \\
        v_{i_1}a_{i_1}^k  &  v_{i_2}a_{i_2}^k  & \cdots & v_{i_{k-1}}a_{i_{k-1}}^k & 1 \\
        \end{array}\right)  
        ~\text{or}~
        M_{k}^{\Rmnum{2}}=\left(\begin{array}{cccc}
        v_{i_1}  & v_{i_2}  &  \cdots & v_{i_k} \\
        v_{i_1}a_{i_1}  & v_{i_2}a_{i_2}  & \cdots & v_{i_k}a_{i_k} \\
        \vdots  & \vdots  & \ddots & \vdots \\
        v_{i_1}a^{k-2}_{i_1}  &  v_{i_2}a^{k-2}_{i_2}  & \cdots & v_{i_k}a^{k-2}_{i_k} \\
        v_{i_1}a_{i_1}^k  &  v_{i_2}a_{i_2}^k  & \cdots & v_{i_k}a_{i_k}^k \\
        \end{array}\right), 
    \end{align*}
    where $1\leq i_1<i_2<\cdots<i_k\leq n$. Then we have
    \begin{align}\label{eq.det_Mk111}
        \det(M_{k}^{\Rmnum{1}}) & = \prod_{j=1}^{k-1} v_{i_j} \prod_{1\leq s<t\leq k-1}(a_{i_t}-a_{i_s})\neq 0,
    \end{align}
    and 
    \begin{align}\label{eq.det_Mk222}
        \det(M_{k}^{\Rmnum{2}}) & = (a_{i_1}+a_{i_2}+\cdots+a_{i_k})\cdot \prod_{j=1}^{k} v_{i_j} \prod_{1\leq s<t\leq k}(a_{i_t}-a_{i_s}). 
    \end{align}

    1)  With \eqref{eq.locality_condition} and \eqref{eq.det_Mk111}, we deduce that 
    $\{i_1,i_2,\ldots,i_{k-1},n+1\}$ cannot be a support of $\ccc^{\perp}$ for any $1\leq i_1<i_2<\cdots<i_k\leq n$.
    Therefore, $$n+1\notin \bigcup_{B\in \mathcal{B}_k\left(\ESGRS_k(\SSS,\mathbf{v},\infty)^{\perp}\right)} B.$$ 
    It turns out from Lemmas \ref{lem.NMDS_cLRC} and \ref{lem.locality_NMDS}.1) that $\ESGRS_k(\SSS,\mathbf{v},\infty)$ has locality $k$. 
    This completes the proof of 1). 

    2) Combining \eqref{eq.locality_condition} and \eqref{eq.det_Mk222}, 
    we see that a minimum weight codeword $\ccc^{\perp}$ has support $\{i_1,i_2,\ldots,i_k\}$ if and only if 
$$
a_{i_1}+a_{i_2}+\cdots+a_{i_k}=0.
$$
Therefore,
\begin{align*}
   \mathcal{B}_k(\ESGRS_k(\SSS,  \mathbf{v},\infty)^{\perp})=
\{\{i_1,i_2,\ldots,i_k\}:~  \{a_{i_1},a_{i_2},\ldots,a_{i_k}\}\subseteq \SSS~{\rm with}~
a_{i_1}+a_{i_2}+\cdots+a_{i_k}=0\}.
\end{align*}

Note also that for any $a_j\in \SSS$, we have
\begin{align}\label{eq.N_recurrence_polished}
N(k,0,\SSS\setminus \{a_j\}) = N(k,0,\SSS)-N(k-1,-a_j,\SSS\setminus  \{a_j\}).
\end{align}
Since $\emptyset\neq \SSS\setminus \{a_j\}\subseteq \SSS$, it follows from \eqref{eq.Nkbs} that 
\begin{align*}
    N(0,-ka_j,\SSS) = N(0,-ka_j,\SSS\setminus \{a_j\}) = \left\{
        \begin{array}{ll}
            1, & \mbox{if}~ ka_j=0, \\
            0, & \mbox{if}~ ka_j\neq 0.
        \end{array}
    \right.
\end{align*}
Then repeated application of \eqref{eq.N_recurrence_polished} yields
\[
N(k,0,\SSS\setminus \{a_j\})=\sum_{i=0}^{k}(-1)^i N\left(k-i,-i a_j,\SSS\right).
\]
Thus, if $\sum_{i=0}^{k}(-1)^i N(k-i,-ia_j,\SSS)> 0$ for any $a_j\in \SSS$, 
then there exists a $k$-subset $\{a_{i_1},a_{i_2},\ldots,a_{i_k}\}\subseteq \SSS\setminus \{a_j\}$ such that
$
a_{i_1}+a_{i_2}+\cdots+a_{i_k}=0.
$
This implies that there is always at least one minimum weight codeword 
$\ccc^{\perp}\in \ESGRS_k(\SSS,\mathbf{v},\infty)^{\perp}$ whose support avoids position $j$, for every $1\leq j\leq n$.

Combining with \eqref{eq.locality_condition} and \eqref{eq.det_Mk222}, we conclude that
\[
\bigcap_{B\in \mathcal{B}_k\left(\ESGRS_k(\SSS,\mathbf{v},\infty)^{\perp}\right)} B=\emptyset.
\]
It then follows from Lemma~\ref{lem.locality_NMDS}.2) that
$\ESGRS_k(\SSS,\mathbf{v},\infty)^{\perp}$ has locality $n-k$.
This completes the proof of 2).
\end{IEEEproof}

\subsection{A related problem on combinatorial designs}

Combinatorial designs and linear codes have close connections \cite{DT2022}. 
Given an $[n,k,d]_q$ linear code $\mathcal{C}$,  
let $\mathcal{P}=\{ 1, 2, \dots, n \}$ and  
let $\mathcal{B}_{d}$ be the set of supports of the codewords of weight $d$ in $\mathcal{C}$. 
If each $t$-subset of $\mathcal{P}$ is contained in exactly $\lambda$ elements of $\mathcal{B}_d$, 
we say that {\em the minimum weight codewords of $\mathcal{C}$ support a $t$-$(n, d, \lambda)$ design}, where $\lambda$ 
is a certain positive integer.  
If $t=n$ or $n=d$, we call such a design \emph{trivial}. 

By removing the last coordinate of an ESGRS code, Han and Zhang~\cite{HZ2024} introduced the class of subcodes of GRS (SGRS) codes.
Note that the authors of~\cite{LZM2024} proved that, in some cases, 
the  minimum weight codewords of SGRS codes support $2$-designs. 
It is also worth noting that many authors have successfully used extended codes 
of linear codes to construct further $t$-designs, while some of the original codes can 
also support $t'$-designs with $t\geq t'$ \cite{DT2022,DL2017,DT2020,SDC2024}.
Motivated by these works, Li \emph{et al.}~\cite{LSZ2025} conducted extensive Magma~\cite{Magma} experiments and observed that 
the  minimum weight codewords of $\ESGRS_k(\SSS,\mathbf{v},\infty)$ do not support nontrivial 
$t$-designs in general, where $t\geq 2$. 
However, no theoretical proof of this phenomenon was provided in that paper. 
Furthermore, they left this as a research problem in~\cite[Section \Rmnum{4}]{LSZ2025}, 
which we rephrase as the following problem.

\begin{problem}{\rm (\!\! \cite[Section \Rmnum{4}]{LSZ2025})}\label{prob.3}
    For given $\SSS$ and $\vvv$, can the minimum weight codewords of $\ESGRS_k(\SSS,\mathbf{v},\infty)$ 
    support nontrivial $t$-designs for some $t\geq 2$?
\end{problem}

Based on Theorem \ref{th.ESGRS_cLRCs}, we {\em disprove} this problem with the following result. 
To this end, we need a useful lemma.

\begin{theorem}\label{th.designs_locality}
    Minimum weight codewords of any ESGRS code cannot support nontrivial $t$-designs for any $t\geq 1$. 
    Therefore, the answer to {\bf Problem}~\ref{prob.3} is negative.
\end{theorem}
\begin{IEEEproof} 
Since MDS codes can only support complete designs, and hence trivial designs, 
it suffices to consider NMDS ESGRS codes by Lemma~\ref{lem.NMDS}. 
Let $\ESGRS_k(\SSS,\mathbf{v},\infty)$ be an $[n+1,k,n-k+1]_q$ NMDS ESGRS code, and assume that 
its minimum weight codewords support a nontrivial $t$-design for some $t\geq 1$. 
Then $\ESGRS_k(\SSS,\mathbf{v},\infty)^\perp$ is an $[n+1,n+1-k,k]_q$ linear code 
whose minimum weight codewords also support a nontrivial $t$-design, 
namely, the complementary design of the original one. 
By Definition~\ref{def.ESGRS_code}, we have $k\geq 3$ and $n-k+1\geq 3$. 
Therefore, both $\ESGRS_k(\SSS,\mathbf{v},\infty)$ and its dual have minimum distance at least $2$. 
Moreover, since every $t$-design with $t\geq 1$ is in particular a $1$-design with alternative parameters, 
the minimum weight codewords of $\ESGRS_k(\SSS,\mathbf{v},\infty)^\perp$ support a nontrivial $1$-design. 
By the definition of locality, an $[n,k]_q$ linear code $\C$ has locality $d(\C^\perp)-1$ if every coordinate 
is contained in the support of some minimum weight codeword of $\C^\perp$. 
This condition is guaranteed when the minimum weight codewords of $\C^\perp$ support a nontrivial $1$-design (see also~\cite[Corollary~3]{TFDTZ2023}). 
Applying this to $\C=\ESGRS_k(\SSS,\mathbf{v},\infty)$, we obtain that the locality of  
$\ESGRS_k(\SSS,\mathbf{v},\infty)$ is 
\[
d\bigl(\ESGRS_k(\SSS,\mathbf{v},\infty)^\perp\bigr)-1=k-1.
\]
This contradicts Theorem~\ref{th.ESGRS_cLRCs}.1). 
Therefore, the minimum weight codewords of $\ESGRS_k(\SSS,\mathbf{v},\infty)$ cannot support any nontrivial $t$-design with $t\geq 1$, yielding a negative answer to {\bf Problem}~\ref{prob.3}. 
This completes the proof.
\end{IEEEproof}

In the following, we provide an example to illustrate Theorems~\ref{th.ESGRS_cLRCs} and~\ref{th.designs_locality}.
In particular, this example shows that the sufficient condition in Theorem~\ref{th.ESGRS_cLRCs}.2) is also necessary in certain cases. 

\begin{example}
    Let $q=11$ and $k=4$. 
    Take $\SSS_1=\{9,7,2,0,8,4,6\}\subseteq \F_{11}$ and $\SSS_2=\{10,1,4,9,2,0,6\}\subseteq \F_{11}$.  
    Let $\ESGRS_4(\SSS_1,{\bf 1}_7,\infty)$ and $\ESGRS_4(\SSS_2,{\bf 1}_7,\infty)$ be the ESGRS codes 
    with respective generator matrices 
    $$\begin{pmatrix}
        1 & 1 & 1 & 1 & 1 & 1 & 1 & 0 \\
        9 & 7 & 2 & 0 & 8 & 4 & 6 & 0 \\
        4 & 5 & 4 & 0 & 9 & 5 & 3 & 0 \\
        5 & 3 & 5 & 0 & 4 & 3 & 9 & 1 \\
    \end{pmatrix},
    $$
    and 
    $$    \begin{pmatrix}
        1 & 1 & 1 & 1 & 1 & 1 & 1 & 0 \\
        10 & 1 & 4 & 9 & 2 & 0 & 6 & 0 \\
        1 & 1 & 5 & 4 & 4 & 0 & 3 & 0 \\
        1 & 1 & 3 & 5 & 5 & 0 & 9 & 1 \\
    \end{pmatrix}.
    $$
    Then we have the following results, 
    where all computations have been carried out or verified using Magma \cite{Magma}.
    \begin{enumerate}
        \item Since $9+7+0+6=0$ in $\SSS_1$ and $1+4+0+6=0$ in $\SSS_2$, 
        by Lemma \ref{lem.NMDS}, both $\ESGRS_4(\SSS_1,{\bf 1}_7,\infty)$ and $\ESGRS_4(\SSS_2,{\bf 1}_7,\infty)$
        are $[8,4,4]_{11}$ NMDS codes.

        \item With Theorem \ref{th.ESGRS_cLRCs}.1), 
        both $\ESGRS_4(\SSS_1,\mathbf{v},\infty)$ and $\ESGRS_4(\SSS_2,\mathbf{v},\infty)$ 
        have locality $4$. In fact, we have   
        $$\{5,8\}\nsubseteq  \bigcup_{B\in \mathcal{B}_4\left(\ESGRS_4(\SSS_1,\mathbf{v},\infty)^{\perp}\right)} B=\{1,2,3,4,6,7\},$$ 
        and 
        $$8\notin  \bigcup_{B\in \mathcal{B}_4\left(\ESGRS_4(\SSS_2,\mathbf{v},\infty)^{\perp}\right)} B=\{1,2,3,4,5,6,7\}.$$ 
        These coincide with the proof of Theorem \ref{th.ESGRS_cLRCs}.1).  

        \item Let 
        $$\Delta_t=\left\{\left(a_j, \sum_{i=0}^{4}(-1)^i N(4-i,-ia_j,\SSS_t)\right):~a_j\in \SSS_t\right\}.$$
        Then 
        $$\Delta_1=\{(9,0),(7,0),(2,1),(0,1),(8,2),(4,1),(6,1)\},$$
        and 
        $$\Delta_2=\{(10,1),(1,1),(4,1),(9,2),(2,1),(0,2),(6,1)\}.$$
        With Theorem \ref{th.ESGRS_cLRCs}.2), we know that 
        $\ESGRS_4(\SSS_2,\mathbf{v},\infty)^{\perp}$ has locality $3$, 
        while $\ESGRS_4(\SSS_1,\mathbf{v},\infty)^{\perp}$ may not have locality $3$. 
        Furthermore, we verify that the locality of $\ESGRS_4(\SSS_1,\mathbf{v},\infty)^{\perp}$ is actually $4$. 

        \item We verify that 
        the minimum weight codewords of both $\ESGRS_4(\SSS_1,\mathbf{v},\infty)$ and 
        $\ESGRS_4(\SSS_2,\mathbf{v},\infty)$ cannot support nontrivial $t$-designs for any $t\geq 1$. 
        This coincides with Theorem \ref{th.designs_locality}. 
    \end{enumerate}
\end{example}

\section{Plotkin-optimal pure qLRCs with larger and more flexible minimum distances}\label{sec.5} 

We now apply the framework established in Theorem \ref{th.pure_qLRCs_from_NMDS} 
to the family of ESGRS codes and construct four families of Plotkin-optimal pure qLRCs. 
Consequently, our results provide specific and affirmative solutions to 
both {\bf Problem} \ref{prob.1} and {\bf Problem} \ref{prob.2}.
To this end, we recall some related codes and two useful lemmas. 

Let $\SSS=\{a_1,a_2,\ldots,a_n\}\subseteq \F_{q}$ and $\mathbf{v}=(v_1,v_2,\ldots,v_n)\in (\F_{q}^*)^n$ be defined as before,  
and let 
$$
\F_q[x]_{k}=\left\{f(x)=\sum_{i=0}^{k-1} f_i x^i:~ f_0, f_1, \ldots,f_{k-1}\in \F_q\right\}. 
$$
Then we have the following definition. 

\begin{definition}{\rm (\!\! \cite{HP2003})}
Given $1\leq k\leq n\leq q$, an $[n,k,n-k+1]_{q}$ {\em generalized Reed-Solomon (GRS) code} is defined as  
$$
\GRS_k(\SSS,\mathbf{v}) =  \Bigl\{(v_1f(a_1),v_2f(a_2),\ldots,v_nf(a_n)):~ f(x)\in {\mathbb{F}_{q}[x]_{k}}\Bigr\}.
$$
By adding a coordinate at the last position of each codeword of a GRS code, 
we get an $[n+1,k,n-k+2]_q$ {\em extended GRS code} 
\begin{align*}
    \GRS_k(\SSS,\mathbf{v},\infty) =  \Bigl\{(v_1f(a_1),v_2f(a_2),\ldots,v_nf(a_n), f_{k-1}): f(x)\in {\mathbb{F}_{q}[x]_{k}}\Bigr\},
\end{align*}
where $f_{k-1}$ is the coefficient of $x^{k-1}$ in $f(x)\in \F_q[x]_{k}$.     
\end{definition}
From \cite{HP2003}, we can write generator matrices of $\GRS_k(\SSS,\mathbf{v})$ 
and $\GRS_k(\SSS,\mathbf{v},\infty)$ as 
\begin{align}\label{eq.GRS.generator matrix}
    G_{\GRS_k(\SSS,\mathbf{v})}=\begin{pmatrix}    
        v_1 & v_2  &  \ldots & v_{n} \\ 
        v_1a_1  &  v_2a_2  & \ldots & v_{n}a_{n} \\
        \vdots  & \vdots  & \ddots & \vdots  \\
        v_1a_1^{k-2}  &  v_2a_2^{k-2}  & \ldots & v_{n}a_{n}^{k-2}  \\
        v_1a_1^{k-1}  &  v_2a_2^{k-1}  & \ldots & v_{n}a_{n}^{k-1}  \\
    \end{pmatrix},
\end{align}
and
\begin{align}\label{eq.EGRS.generator matrix}
    G_{\GRS_k(\SSS,\mathbf{v},\infty)}=\begin{pmatrix}    
        v_1 & v_2 &  \ldots & v_{n} & 0 \\ 
        v_1a_1  &  v_2a_2  & \ldots & v_{n}a_{n} & 0 \\
        \vdots  & \vdots  & \ddots & \vdots & \vdots \\
        v_1a_1^{k-2}  &  v_2a_2^{k-2}  & \ldots & v_{n}a_{n}^{k-2} & 0 \\
        v_1a_1^{k-1}  &  v_2a_2^{k-1}  & \ldots & v_{n}a_{n}^{k-1} & 1 \\
    \end{pmatrix},
\end{align}
respectively.

\begin{lemma}{\rm (\!\! \cite[Theorem 1.1]{li2008subset})}\label{lem.ssp_01}
    Let $q^2\geq 4$ and let $\mathcal{D}\subseteq \F_{q^2}$ be a subset of cardinality $n$. 
    For any $1\leq k\leq n\leq q^2-2$ and any $b\in \F_{q^2}$, we have  
    $$
    \left|N(k,b,\mathcal{D})-\frac{1}{q^2}\binom{n}{k}\right|\leq \frac{q^2-1}{q^2} 
    \binom{k+q^2-n-2}{q^2-n-2} \binom{\frac{q^2}{p}}{\lfloor \frac{k}{p} \rfloor}.
    $$
\end{lemma}

\begin{lemma}{\rm (\!\! \cite[Corollaries 2.7 and 2.8]{li2008subset})}\label{lem.ssp_additive_subgroup}
    Let $q^2\geq 9$ and $b\in \F_{q^2}$. 
    If $p=2$ and $2\leq k\leq q^2-3$, or $p$ is odd and $3\leq k\leq q^2-4$, 
    then both $N(k,b,\F_{q^2})>0$ and $N(k,b,\F_{q^2}^*)>0$.  
\end{lemma}

In the following, we obtain the first family of Plotkin-optimal pure qLRCs with larger and more flexible minimum distances.

\begin{theorem}\label{th.qlrc111}
    Let $q=p^h\geq 5$ be a prime power with $p\geq 3$. 
    For any $\left\lceil \frac{2p}{p-2} \right\rceil\leq k\leq q-2$, 
    there exists a family of Plotkin-optimal pure 
    $$\Dsb{q^2,q^2-2k,k}_q$$ qLRCs with locality $r=q^2-k-1$. 
    Furthermore, these qLRCs are also optimal with respect to the pure Singleton-like bound in \eqref{eq.pure_Singleton} 
    and the pure Griesmer-like bound in \eqref{eq.pure_Griesmer}, simultaneously.
\end{theorem}
\begin{IEEEproof}
    Let $\omega$ be a primitive element of $\F_{q^2}$ and write 
    $$
        \F_{q^2}=\left\{0,1,\omega,\omega^2,\ldots,\omega^{q^2-2}\right\}
    $$
    with the prescribed order. 
    Note that Jin $et~al.$ showed in \cite[Lemma 2.2.ii)]{JLLX2010} that for any $k+1\leq q-1$, 
    there exists $\mathbf{v}=(\beta,1,1,\ldots,1)\in (\F_{q^2}^*)^{q^2}$ with $\beta^{q+1}=1$ such that 
    $\GRS_{k+1}(\F_{q^2},\mathbf{v})$ is Hermitian self-orthogonal. 
    Taking $\beta=1$, it then turns out from \eqref{eq.GRS.generator matrix} and \eqref{eq.EGRS.generator matrix} 
    that 
    $$\GRS_{k+1}(\F_{q^2},\mathbf{1})=\EGRS_{k+1}(\SSS',\mathbf{v}',\infty)$$ 
    is Hermitian self-orthogonal, 
    where $$\SSS'=\{1,\omega^{q^2-2},\omega^{q^2-3},\ldots,\omega\}$$ and
    $$\vvv'=(1,\omega^{k},\omega^{2k},\ldots,\omega^{(q^2-2)k})\in (\F_{q^2}^*)^{q^2-1}.$$ 
    Combining \eqref{eq.ETGRS.generator_matrix} and \eqref{eq.EGRS.generator matrix}, we also know that 
    $\ESGRS_{k}(\SSS',\mathbf{v}',\infty)$ is a subcode of the 
    extended GRS code $\EGRS_{k+1}(\SSS',\mathbf{v}',\infty)$. It then implies that 
\begin{align}\label{eq.nested}
    \begin{split}
  \ESGRS_{k}(\SSS',\vvv',\infty)  
 & \subseteq   \EGRS_{k+1}(\SSS',\vvv',\infty)  \\
 & \subseteq   \EGRS_{k+1}(\SSS',\vvv',\infty)^{\perp_{\rm H}}  
 \subseteq   \ESGRS_{k}(\SSS',\vvv',\infty)^{\perp_{\rm H}}.
     \end{split}
\end{align}
That is to say $\ESGRS_{k}(\SSS',\mathbf{v}',\infty)^{\perp_{\rm H}}$ is Hermitian dual-containing 
for any $3\leq k\leq q-2$ and $q\geq 5$. 

We now consider the parameters and localities of $\ESGRS_{k}(\SSS',\mathbf{v}',\infty)^{\perp_{\rm H}}$. 
Since the subset sum problem does not depend on the order of elements in the set, 
it follows from Lemma \ref{lem.ssp_additive_subgroup} that 
$$
N(k,0,\SSS')=N(k,0,\F_{q^2}^*)>0.   
$$
Then, applying Lemma~\ref{lem.NMDS}, we conclude that 
$\ESGRS_{k}(\SSS',\mathbf{v}',\infty)^{\perp_{\rm H}}$ is a $[q^2,q^2-k,k]_{q^2}$ NMDS code. 
For the locality, we need to verify that 
\begin{align}\label{eq.ssp11}
    \sum_{i=0}^{k}(-1)^i N(k-i,-ia_j,\SSS')>0
\end{align}
holds for any $a_j\in \F_{q^2}^*$. 
Since $|\F_{q^2}\setminus \{0,a_j\}|=q^2-2$, we deduce from Lemma \ref{lem.ssp_01} that 
\begin{align*}
     \sum_{i=0}^{k}(-1)^i N(k-i,-ia_j,\SSS') 
 =  N(k,0,\F_{q^2}\setminus \{0,a_j\})  
\geq  \frac{1}{q^2}\binom{q^2-2}{k}- \frac{q^2-1}{q^2} 
    \binom{\frac{q^2}{p}}{\lfloor \frac{k}{p} \rfloor}.
\end{align*}
Note that $p\geq 3$ and $\left\lceil \frac{2p}{p-2} \right\rceil\leq k\leq q-2<q-1<q^2-1$, 
then $q^2-k-1>0$, $\frac{q-1}{k}>1$, and $k-2\lfloor \frac{k}{p} \rfloor-2\geq 0$.
By the definitions of combinatorial numbers and the value ranges of each involving parameter, 
we further obtain that 
\begin{align*}
    \frac{\binom{q^2-2}{k}}{(q^2-1)\binom{\frac{q^2}{p}}{\lfloor \frac{k}{p} \rfloor}} 
    & = \frac{1}{q^2-1}\cdot \frac{\lfloor \frac{k}{p} \rfloor!}{k!}\cdot \frac{\prod_{i=0}^{k-1}(q^2-2-i)}{\prod_{j=0}^{\lfloor \frac{k}{p} \rfloor-1}(\frac{q^2}{p}-1-j)} \\ 
    & \geq \frac{1}{q^2}\cdot \frac{1}{k^{k-\lfloor \frac{k}{p} \rfloor}}\cdot \frac{(q^2-k-1)^k}{\left(\frac{q^2}{p} \right)^{\lfloor \frac{k}{p} \rfloor}} \\ 
    & \geq  \frac{1}{q^2}\cdot \frac{1}{k^{k-\lfloor \frac{k}{p} \rfloor}}\cdot \frac{(q^2-q)^k}{q^{2\lfloor \frac{k}{p} \rfloor}}\cdot p^{\lfloor \frac{k}{p} \rfloor} \\ 
    & = q^{k-2\lfloor \frac{k}{p} \rfloor-2} \cdot \left(\frac{q-1}{k}\right)^k\cdot (pk)^{\lfloor \frac{k}{p} \rfloor} \\
    & > 1,
\end{align*} 
implying $\frac{1}{q^2}\binom{q^2-2}{k}> 
\frac{q^2-1}{q^2} \binom{\frac{q^2}{p}}{\lfloor \frac{k}{p} \rfloor}$, 
and hence \eqref{eq.ssp11} indeed holds. 
It then turns out from Theorem \ref{th.ESGRS_cLRCs}.2) that 
$\ESGRS_{k}(\SSS',\vvv',\infty)^{\perp}$, and hence $\ESGRS_{k}(\SSS',\vvv',\infty)^{\perp_{\rm H}}$ has locality $r=q^2 - k-1$. 

Finally, applying Lemma \ref{lem.Hermitian_qLRC} and Theorem \ref{th.pure_qLRCs_from_NMDS} to 
$\ESGRS_{k}(\SSS',\vvv',\infty)^{\perp_{\rm H}}$ immediately yields the desired result. 
This completes the proof.
\end{IEEEproof}


\begin{remark}\label{rem.second-family} 
{We include the following remark. 
The first part gives further explanations of the parameters in Theorem~\ref{th.qlrc111}, 
while the second part presents another family of Plotkin-optimal pure qLRCs that complements Theorem~\ref{th.qlrc111}.}
    \begin{enumerate} 
        \item It can be easily verified that 
    \begin{align*}
        \left\lceil \frac{2p}{p-2} \right\rceil
        =2+\left\lceil \frac{4}{p-2} \right\rceil = \left\{\begin{array}{ll}
            6, & \mbox{if}~ p=3, \\
            4, & \mbox{if}~ p=5, \\ 
            3, & \mbox{if}~ p\geq 7.
        \end{array}\right.
    \end{align*}
    According to the proof of Theorem \ref{th.qlrc111}, there is no loss on the possible 
    values of $k$ such that $\ESGRS_{k}(\SSS',\vvv',\infty)^{\perp_{\rm H}}$ is Hermitian dual-containing
    and has locality $r=q^2-k-1$.  

    \item {To ensure that the term $\lceil \frac{2p}{p-2} \rceil$ is well defined, 
we exclude the case $p=2$ in Theorem~\ref{th.qlrc111}. 
Note that Li \emph{et al.}~\cite{LLLLL2025} employed Hermitian dual-containing NMDS SGRS codes supporting $3$-designs to construct a family of 
\[
\Dsb{4^m,4^m-2k,k}_{2^m}
\]
qLRCs with locality $r=4^m-k-1$, where $k$ is even, $3\leq k\leq 2^m-2$, and $m\geq 3$. 
Recall that this is the only known family of optimal pure qLRCs with both large lengths and minimum distances, 
and the optimality established so far is with respect to the pure Singleton-like bound in \eqref{eq.pure_Singleton}. 
By applying Theorem~\ref{th.pure_qLRCs_from_NMDS}, we further conclude that these qLRCs are also simultaneously optimal with respect to 
the pure Griesmer-like bound in \eqref{eq.pure_Griesmer} and the pure Plotkin-like bound in \eqref{eq.pure_Plotkin}. 
Therefore, although this family was originally constructed in~\cite{LLLLL2025}, our framework identifies it 
for the first time as a family of Plotkin-optimal pure qLRCs. 
In this sense, we regard it as the second new family of Plotkin-optimal pure qLRCs obtained from our framework, 
complementing Theorem~\ref{th.qlrc111} for the case $p=2$. 
In view of the discussion in 1) above, we also emphasize that our first family of Plotkin-optimal pure qLRCs allows more flexible choices 
of the minimum distance compared to the second family, since it removes the parity restriction $2\mid k$.
}
\end{enumerate}
\end{remark}


In Theorem~\ref{th.qlrc111}, the case $p=2$ is excluded, 
while Remark~\ref{rem.second-family}.2) provides a complementary family for this case. 
We now show that, by allowing the length to be slightly smaller than $q^2$ and by making a minor reduction in the minimum distance, 
one can obtain Plotkin-optimal pure qLRCs over arbitrary finite fields. 
This gives our third family of Plotkin-optimal pure qLRCs.

\begin{theorem}\label{th.qlrc222}
    Let $q=p^h\geq 7$ be a prime power. 
    Let $\ell$ and $k$ be two integers satisfying $1\leq \ell\leq q-5$ and $3 \leq k\leq q-\ell-2.$ 
    Let $[k]_p$ be the minimum nonnegative integer congruent to $k$ modulo $p$.
    If 
    \begin{align}\label{eq.th2cond}
    \ell \leq 2k-(p+1)\left\lfloor \frac{k}{p} \right\rfloor+\log_q \left(\frac{p^{\left\lfloor \frac{k}{p} \right\rfloor}}{2^k}\right)-[k]_p-2,        
    \end{align}
    then there exists a family of Plotkin-optimal pure 
    $$\Dsb{q^2-\ell,q^2-2k-\ell,k}_q$$ qLRCs with locality $r=q^2-k-\ell-1$. 
    Furthermore, these qLRCs are also optimal with respect to the pure Singleton-like bound in \eqref{eq.pure_Singleton} 
    and the pure Griesmer-like bound in \eqref{eq.pure_Griesmer}, simultaneously.
\end{theorem}
\begin{IEEEproof}
    According to \cite[Theorem 6]{LXW2008}, there exists a $[q^2-\ell,k+1,q^2-\ell-k]_{q^2}$ Hermitian self-orthogonal 
    GRS code $\GRS_{k+1}(\SSS, \mathbf{v})$ for any $1\leq \ell\leq q-2$ and $1\leq k+1\leq q-\ell-1$. 
    Since $q^2-\ell\leq q^2-1$, it then follows from \cite[Remark 6.2]{P1992} that 
    $\GRS_{k+1}(\SSS, \mathbf{v})=\EGRS_{k+1}(\SSS', \mathbf{v}',\infty)$ for some 
    $\SSS'\subseteq \F_{q^2}$ with $|\SSS'|=q^2-\ell-1$ and 
    $\mathbf{v}'\in (\F_{q^2}^*)^{q^2-\ell-1}$. 
    With an argument similar to \eqref{eq.nested}, we immediately obtain that 
    $\ESGRS_{k}(\SSS',\mathbf{v}',\infty)^{\perp_{\rm H}}$ is Hermitian dual-containing 
    for any $3 \leq k\leq q-\ell-2$ and $1\leq \ell\leq q-5$.

We next show that $\ESGRS_{k}(\SSS',\mathbf{v}',\infty)^{\perp_{\rm H}}$ is a $[q^2-\ell,q^2-k-\ell,k]_{q^2}$ 
NMDS code with locality $r=q^2-k-\ell-1$. 
From Lemma~\ref{lem.NMDS} and Theorem \ref{th.ESGRS_cLRCs}.2), 
it suffices to verify that 
$$
N(k,0,\SSS')>0~{\rm and}~
\sum_{i=0}^{k}(-1)^i N(k-i,-ia_j,\SSS')=N(k,0,\SSS'\setminus \{a_j\})>0
$$ 
hold for any $a_j\in \SSS'$.
Since $|\SSS'|=q^2-\ell-1$, we first evaluate $\binom{q^2-\ell-1}{k}$ 
and $(q^2-1)\binom{k+\ell-1}{\ell-1}\binom{\frac{q^2}{p}}{\left\lfloor \frac{k}{p} \right\rfloor}$. 
Since $k\leq q-\ell-2$ and $q\geq 7$, we have $q^2-\ell-k\geq q^2-q+2>\frac{q^2}{2}$ and $k+\ell-1\leq q-3<q$, 
implying that 
\begin{align}\label{eq.th2111}
    \binom{q^2-\ell-1}{k}>\frac{q^{2k}}{k! 2^k} 
\end{align}
and 
\begin{align}\label{eq.th2222}
    (q^2-1)\binom{k+\ell-1}{\ell-1}\binom{\frac{q^2}{p}}{\left\lfloor \frac{k}{p} \right\rfloor} 
    < \frac{q^{2\left\lfloor \frac{k}{p} \right\rfloor+\ell+1}}{\left\lfloor \frac{k}{p} \right\rfloor!p^{\left\lfloor \frac{k}{p} \right\rfloor}}.
\end{align}
Combining \eqref{eq.th2111} and \eqref{eq.th2222}, we conclude from Lemma \ref{lem.ssp_01} that 
if 
\begin{align}\label{eq.th2333}
    \frac{k!}{\left\lfloor \frac{k}{p} \right\rfloor!}<k^{k-\left\lfloor \frac{k}{p} \right\rfloor} \leq q^{2k-2\left\lfloor \frac{k}{p} \right\rfloor-\ell-1} \cdot \frac{p^{\left\lfloor \frac{k}{p} \right\rfloor}}{2^k},
\end{align}
then $N(k,0,\SSS')>0$. 
Note that $k-\left\lfloor \frac{k}{p} \right\rfloor=(p-1)\left\lfloor \frac{k}{p} \right\rfloor+[k]_p$ and $k<q$, 
then \eqref{eq.th2333} holds provided that 
$$\ell \leq 2k-(p+1)\left\lfloor \frac{k}{p} \right\rfloor+\log_q \left(\frac{p^{\left\lfloor \frac{k}{p} \right\rfloor}}{2^k}\right)-[k]_p-1.$$ 
Since $|\SSS'\setminus \{a_j\}|=q^2-\ell-2$, it turns out from 
an argument similar to the above that $N(k,0,\SSS'\setminus \{a_j\})>0$ holds for any $a_j\in \SSS'$ 
if $$\ell \leq 2k-(p+1)\left\lfloor \frac{k}{p} \right\rfloor+\log_q \left(\frac{p^{\left\lfloor \frac{k}{p} \right\rfloor}}{2^k}\right)-[k]_p-2.$$  
In summary, the condition required in \eqref{eq.th2cond} ensures that 
$\ESGRS_{k}(\SSS',\mathbf{v}',\infty)^{\perp_{\rm H}}$ is a $[q^2-\ell,q^2-k-\ell,k]_{q^2}$ 
NMDS code with locality $r=q^2 - k-\ell-1$.

Again, applying Lemma \ref{lem.Hermitian_qLRC} and Theorem \ref{th.pure_qLRCs_from_NMDS} to 
$\ESGRS_{k}(\SSS',\vvv',\infty)^{\perp_{\rm H}}$ immediately yields the desired result. 
This completes the proof.
\end{IEEEproof}

Finally, we present the fourth family of $q$-ary Plotkin-optimal pure qLRCs of length $q^2+1$.

\begin{theorem}\label{th.qlrc333}
    Let $q=p^h\geq 4$ be any prime power. 
    For any $3\leq k\leq q-1$ and $k\neq q-2$, there exists a family of 
    Plotkin-optimal pure 
    $$\Dsb{q^2+1,q^2-2k+1,k}_q$$ qLRCs with locality $r=q^2-k$. 
    Furthermore, these qLRCs are also optimal with respect to the pure Singleton-like bound in \eqref{eq.pure_Singleton} 
    and the pure Griesmer-like bound in \eqref{eq.pure_Griesmer}, simultaneously.
\end{theorem}
\begin{IEEEproof}
    Consider the full length $[q^2+1,k]_{q^2}$ ESGRS code $\ESGRS_{k}(\F_{q^2},\mathbf{v},\infty)$ 
    with $\mathbf{v}=(v_1,v_2,\ldots,v_{q^2})\in (\F_{q^2}^*)^{q^2}$. 
From \cite[Theorem 4]{B2021}, there exists $\vvv'\in (\F_{q^2}^*)^{q^2+1}$ such that 
$\GRS_{k+1}(\F_{q^2},\vvv',\infty)$ is 
Hermitian self-orthogonal for any $k+1\leq q$ and $k+1\neq q-1$. 
Again, applying an argument similar to that in \eqref{eq.nested}, we deduce that 
$\ESGRS_{k}(\F_{q^2},\vvv',\infty)^{\perp_{\rm H}}$ is Hermitian dual-containing 
for any $3\leq k\leq q-1$, $k\neq q-2$, and $q\geq 4$. 

With Lemma \ref{lem.ssp_additive_subgroup}, it is straightforward to verify that 
$$N(k,0,\F_{q^2})>0$$ 
and 
\begin{align*}
     \sum_{i=0}^{k}(-1)^i N(k-i,-ia_j,\F_{q^2}) 
=  N(k,0,\F_{q^2}\setminus \{a_j\}) 
=  N(k,-k a_j,\F_{q^2}^*)
>  0. 
\end{align*}
It then follows from Lemma \ref{lem.NMDS} and Theorem \ref{th.ESGRS_cLRCs}.2) that 
$\ESGRS_{k}(\F_{q^2},\vvv',\infty)^{\perp_{\rm H}}$ is a $[q^2+1,q^2+1-k, k]_{q^2}$ 
NMDS code with locality $r=q^2 - k$. 

Therefore, the desired result follows by applying Lemma \ref{lem.Hermitian_qLRC} and Theorem \ref{th.pure_qLRCs_from_NMDS} to 
$\ESGRS_{k}(\F_{q^2},\vvv',\infty)^{\perp_{\rm H}}$. 
This completes the proof.
\end{IEEEproof}

\begin{remark}\label{rem.comparasion}
In the possible ranges of $k$ in Theorems \ref{th.qlrc111}, \ref{th.qlrc222}, and \ref{th.qlrc333}, as well as Remark \ref{rem.second-family}, 
we always have $(r+1)\nmid q^2$, $q^2-\ell$, or $q^2+1$. 
Moreover, in conjunction with Remark~\ref{rem.CSS_bounds}, it can be easily verified that the optimal pure qLRCs 
listed in Table~\ref{tab:qLRCs} are, in general, not Plotkin-optimal. 
In this sense, our framework and constructions not only provide the first four families of optimal qLRCs with parameters 
attaining the pure Plotkin-like bound, but also admit larger and more flexible minimum distances 
than those of existing constructions. 
As we have emphasized before, larger and more flexible minimum distances provide greater freedom 
in balancing the parameters required for both local and global quantum recovery.  
\end{remark}


\begin{example}
    Let $q=9$. We immediately obtain eight Plotkin-optimal pure qLRCs from Theorems \ref{th.qlrc111}, \ref{th.qlrc222}, and \ref{th.qlrc333}. 
    Note that the second family of Plotkin-optimal pure qLRCs in Remark \ref{rem.second-family} does not exist in this setting 
    and the value of $\ell$ related to Theorem \ref{th.qlrc222} is fixed to $1$. 
    Table \ref{tab:example_qLRCs} summarizes their explicit parameters. 
    Note that if larger $q$ is taken, more Plotkin-optimal pure qLRCs with larger minimum distances can be obtained similarly.

\begin{table*}[ht]
\centering
\caption{Some Plotkin-optimal pure qLRCs over $\F_9$}
\label{tab:example_qLRCs}
\setlength{\tabcolsep}{4pt}
\renewcommand{\arraystretch}{1.45}
\setlength{\extrarowheight}{2pt}
\begin{tabular}{c|c|c|c||c|c|c|c}
\toprule
No. & $\dsb{n,\kappa,\delta}_9$ & $r$ & Reference
& No. & $\dsb{n,\kappa,\delta}_9$ & $r$ & Reference\\ 
\midrule

1 & $\Dsb{80,68,6}_9$ & $73$ & Theorem~\ref{th.qlrc222}  
& 2 & $\Dsb{81,69,6}_9$ & $74$ & Theorem~\ref{th.qlrc111}  
\\ \hline 

3 & $\Dsb{81,67,7}_9$ & $73$ & Theorem~\ref{th.qlrc111}  
& 4 & $\Dsb{82,76,3}_9$ & $78$ & Theorem~\ref{th.qlrc333}  
\\ \hline

5 & $\Dsb{82,74,4}_9$ & $77$ & Theorem~\ref{th.qlrc333}  
& 6 & $\Dsb{82,72,5}_9$ & $76$ & Theorem~\ref{th.qlrc333}  
\\ \hline

7 & $\Dsb{82,70,6}_9$ & $75$ & Theorem~\ref{th.qlrc333}  
& 8 & $\Dsb{82,66,8}_9$ & $73$ & Theorem~\ref{th.qlrc333}  
\\

\bottomrule
\end{tabular}
\end{table*}




\end{example}



\section{Concluding remarks}\label{sec.6}
This paper aimed to address {\bf Problem} \ref{prob.1} and {\bf Problem} \ref{prob.2} 
in a unified manner by using Hermitian dual-containing ESGRS codes. 
We first developed a general framework for constructing optimal pure qLRCs that attain the pure Plotkin-like bound in \eqref{eq.pure_Plotkin}, 
based on Hermitian dual-containing NMDS codes in Theorem \ref{th.pure_qLRCs_from_NMDS}. 

Focusing on the specific family of NMDS ESGRS codes, 
we respectively determined the localities of ESGRS codes and their duals by analyzing the existence of 
nonsingular submatrices in the generator matrices of ESGRS codes and certain subset sum problems 
over finite fields in Theorem \ref{th.ESGRS_cLRCs}. 
Building on this result, we also resolved {\bf Open Problem} \ref{prob.3}, originally proposed in \cite{LSZ2025}, 
{\em in the negative}, as established in Theorem \ref{th.designs_locality}.

Applying this framework to ESGRS codes, we explicitly obtained four new families of Plotkin-optimal pure qLRCs in 
Theorems \ref{th.qlrc111}, \ref{th.qlrc222}, and \ref{th.qlrc333}, as well as Remark \ref{rem.second-family}, 
whose parameters were summarized in Table \ref{tab:our_qLRCs}.
These constructions also achieve optimality simultaneously with respect to the pure Singleton-like bound in \eqref{eq.pure_Singleton}  
and the pure Griesmer-like bound in \eqref{eq.pure_Griesmer}. 
Compared with the known qLRCs with large lengths listed in Table~\ref{tab:qLRCs}, 
our constructions allow larger and more flexible minimum distances, thereby providing 
greater freedom in balancing the parameters required for both local and global quantum recovery \cite{LLLLL2025}.  
Consequently, we give affirmative answers to {\bf Problem} \ref{prob.1} and {\bf Problem} \ref{prob.2}. 
In particular, our results settle a problem proposed by Cao and Zhou in~\cite[Section \Rmnum{7}]{CZ2025-TIT}. 

There are also some interesting questions left for future research. 
We invite the readers to study them together with us. 
\begin{enumerate}

    \item Can we construct more optimal pure qLRCs with larger lengths, 
    particularly exceeding $q^2+c$ for some $c$ depending linearly on $q$? 
    Similar to the MDS conjecture for linear codes, it is conjectured in \cite{DL1995,LR2015} that 
    the maximum length of an NMDS code over $\F_{q^2}$ is at most $2q^2+2$. 
    It is also known that elliptic curve NMDS codes can achieve lengths up to $q^2+2q+1$ \cite{TV1991,W1969}. 
    However, it remains unclear whether such NMDS codes are always Hermitian dual-containing and whether they have nontrivial localities. 
    Once NMDS codes satisfying these conditions are identified, Theorem \ref{th.pure_qLRCs_from_NMDS} can be directly applied to 
    obtain new optimal pure qLRCs with larger lengths.

    \item Would pure qLRCs constructed using Hermitian dual-containing $\ell$-MDS codes with $\ell\geq 2$ exhibit better performance? 
    This is a natural research problem motivated by Remark \ref{rem.lmds}. 
\end{enumerate}




\end{sloppypar}
\end{document}